\documentclass[aps,pra,twocolumn,showpacs,preprintnumbers,letterpaper]{revtex4}
\usepackage{amsmath, amsthm, amssymb}
\usepackage{graphicx}

\DeclareMathOperator{\tr}{tr}
\DeclareMathOperator{\diag}{diag}                       
\newcommand{\sop}[1]{\mathcal{#1}} 	   

\newcommand{\ket}[1]{ |#1\rangle}
\newcommand{\bra}[1]{ \langle #1|}

\newcommand{\sub}[1]{\ensuremath{_{_{\text{#1}}}}}

\newcommand{\subB}{\ensuremath{_{_{\text B}}}}
\newcommand{\subS}{\ensuremath{_{_{\text S}}}}
\newcommand{\subSB}{\ensuremath{_{_{\text{SB}}}}}

\newcommand{\sqrtPiB}{\ensuremath{\pi_{_{\text B}}^{+\frac{1}{2}} } }

\newcommand{\sqrtPiS}{\ensuremath{\pi_{_{\text S}}^{+\frac{1}{2}} } }
\newcommand{\nsqrtPiS}{\ensuremath{\pi_{_{\text S}}^{-\frac{1}{2}} } }

\begin{document}

\title{Quantum Operation Time Reversal}
\author{Gavin E. Crooks }
\email{gecrooks@lbl.gov}
\affiliation{Physical Biosciences Division, Lawrence Berkeley National Laboratory, Berkeley, California 94720}
\date{\today}

\begin{abstract}
The dynamics of an open quantum system can be described by a quantum operation, a linear, complete positive map of operators. Here, I exhibit a compact expression for the time reversal of a quantum operation, which is closely analogous to the time reversal of a classical Markov transition matrix. Since open quantum dynamics are stochastic, and not, in general, deterministic, the time reversal is not, in general, an inversion of the dynamics. Rather, the system relaxes towards equilibrium in both the forward and reverse time directions.  The probability of a quantum trajectory and the conjugate, time reversed trajectory are related by the heat exchanged with the environment. 
\end{abstract}
	
\pacs{05.30.Ch, 05.70.Ln}
\preprint{LBNL-62800}
\maketitle

\section*{Introduction}

Consider a  sequence of states sampled from a classical, homogeneous, steady-state, Markov chain~\cite{Kemeny1976,Norris1997}. For example, 
 \[
 A,\, B,\, C,\, A,\, A,\, B,\, C,\, A,\, B,\, C,\, A\,
 \] 
 where the three states are labeled ($A,B,C$) and time is read from left to right.
The reversed sequence of states, reading from right to left, is also Markovian, homogeneous in time and has the same equilibrium probability distribution.  Moreover, the probability of the transition
$i\rightarrow j$ in the forward chain will be the same as the
probability of the opposite  transition $j\rightarrow i$ in the time
reversed chain. Classical Markov dynamics are equipped with a natural time-reversal operation. 

The quantum mechanical generalization of a Markov transition matrix is a quantum operation, a linear, trace preserving, complete positive (TCP) map of operators~\cite{Kraus1983,Schumacher1996,Caves1999,Nielsen2000}. 
These superoperators can describe a wide range of dynamics, 
including the pure quantum dynamics of an isolated system, the mixed quantum-classical dynamics of
a system interacting with the environment,  and the disturbance induced by measurements of the system, either projection into a subspace due to a von Neumann measurement, or a more general positive operator valued measurement (POVM)~\cite{Nielsen2000}. 

In this note, we consider the generalization of classical Markov chain time reversal to quantum operations. In a quantum system, we cannot meaningfully consider the chain of states. Instead, we consider the time-reversal of the chain of interactions between the system and environment.  The time-reversal invariance of an isolated system is broken by coupling the system to the environment, and the magnitude of this symmetry breaking is related to the environmental entropy change. This naturally leads to an exposition of the work fluctuation theorem for dissipative quantum dynamics.


\section*{Background: Markov chain time reversal}

 Let $M$ be the classical Markov transition matrix of a forward chain of states (i.e.
$M_{ji}$ is the probability of moving from state $i$ to state $j$),
$\widetilde{M}$ the transition matrix of the reversed chain and
$p^{\circ}$ the equilibrium probability distribution of both chains (i.e.
$M p^{\circ} = \widetilde{M} p^{\circ} = p^{\circ}$).  Since 
the probability of the transition
$i\rightarrow j$ in the forward chain will be the same as the
probability of the opposite  transition $j\rightarrow i$ in the time
reversed chain, it follows that the forward and reversed chain transition matrices are related by
	\begin{equation}
	    \widetilde{M}_{ij} \,p^{\circ}_{j}=  M_{ji} \,p^{\circ}_{i}\qquad \mbox{for all} \quad 
	    i,j \, .
	\end{equation}
Note that in much of the Markov chain literature (e.g.~\cite{Norris1997})  the transition matrix is the transpose of the matrix defined here. Also, we have avoided the conventional notation for the time reversed matrix $\hat{M}$ since this may cause unfortunate confusion in the context of quantum dynamics.

In matrix notation this time reversal operation can be conveniently expressed as 
	\begin{equation}
			\widetilde{M} =\diag(p^{\circ}) M^{T} \diag(p^{\circ})^{-1} .
	\end{equation}
 Here, $\mbox{diag}(p^{\circ})$ indicates a matrix whose diagonal 
elements are given by the vector $p^{\circ}$. $\widetilde{M}$ is referred to as 
the reversal of $M$~\cite{Norris1997}, or as the dual of $M$~\cite{Kemeny1976}.
A transition matrix $M$ is \emph{balanced} with respect to a probability distribution $p^{\circ}$ if 
$Mp^{\circ} = p^{\circ}$, and \emph{detailed balanced} if the matrix is time-reversal invariant, $\widetilde{M} = M$.

\section*{Background: Quantum operations}

The dynamics of an open quantum system can be represented as the closed system unitary
dynamics of the system coupled to an extended environment, followed by a measurement of the environment.
\begin{equation}
\rho' = \sop{S} \rho = \tr\sub{E} U\sub{SE} [ \rho \oplus \rho\sub{E}] U\sub{SE}^{\dagger}
\end{equation}
Here $\rho$ and $\rho'$ are the initial and final density matrices of the system, $\rho\sub{E}$ is the initial density matrix of the environment, $U\sub{SE}$ is the unitary operator representing the time evolution of the combined system over some time interval, and $\tr\sub{E}$ is a partial trace over the environment Hilbert space.  The superoperator $\sop{S}$ is a quantum operation, a linear, trace preserving, complete positive 
(TCP) map of operators~\cite{Kraus1983,Schumacher1996,Caves1999,Nielsen2000}. 

Any complete map of positive operators has an operator-sum (or Kraus) 
	representation,
\begin{equation}
	      \rho'  = \sop{S} \rho = \sum_{\alpha} A_{\alpha} \rho A^{\dagger}_{\alpha}
	     \label{quantumop}
\end{equation}
 Conversely, any operator-sum represents a  complete, 
positive superoperator.  The collection $\{A_{\alpha}\}$ are known as Kraus operators.

The requirement that the quantum operation conserved the density matrix
trace can be compactly written as 
 \begin{equation}
 \sop{S}^{\times} I =  \sum_{\alpha}  A^{\dagger}_{\alpha} A_{\alpha} =I
\; .
\end{equation}
Here,  $\sop{S}^{\times}$ is the superoperator adjoint of $\sop{S}$,
the unique
superoperator such that $\langle \sop{S}A, B\rangle = \langle A,
\sop{S}^{\times}B\rangle$, where $\langle A, B\rangle$ is the
Hilbert-Schmit inner product $\tr A^{\dagger} B$.  In the
operator sum representation the superoperator adjoint is performed by
taking the adjoints of the corresponding Kraus operators~\cite{Caves1999}.
\begin{equation}
\sop{S} \rho = \sum_{\alpha} A_{\alpha} \rho 
	A_{\alpha}^{\dagger}, \qquad
	\sop{S}^{\times} \rho = \sum_{\alpha} A_{\alpha}^{\dagger} \rho A_{\alpha}
\end{equation}

Each Kraus operator of a
TCP map represents a particular interaction with the
environment that an external observer could, in principle, measure
and record without further disturbing the dynamics of the system.  The probability of observing the $\alpha$th Kraus interaction is
\begin{equation}
p_{\alpha}  = \tr A_{\alpha}  \rho A^{\dagger}_{\alpha} 
\label{interaction}
\end{equation}
 and the state of the system after this interaction is
\begin{equation}
\rho'_{\alpha} = 
\frac{A_{\alpha}  \rho A^{\dagger}_{\alpha} }{\tr A_{\alpha}  \rho A^{\dagger}_{\alpha} } \,.
\end{equation}
The overall effect of the dynamics, averaging over different interactions, is the full quantum operation, Eq.~(\ref{quantumop}).

In the limit of small time interval we obtain a continuous time quantum Markovian dynamic, 
\begin{equation}
\rho(t) = \exp\left(\int_s^t \sop{L}(\tau) d\tau \right) \rho(s)
\label{Lindbladian}
\end{equation}
where $\sop{L}$ is the Lindbladian 
superoperator~\cite{Nielsen2000}.


\section*{Quantum operation time reversal}
We will now consider the time reversal of a quantum operation. Since we
cannot observe a sequence of states for the quantum dynamics (at least, not without measuring,
and therefore disturbing the system) we instead focus on the sequence of transitions.  
Each operator of a Kraus operator-sum represents a particular interaction with the
environment that an external observer could, in principle, measure
and record.  We can therefore define
the dynamical history by the observed sequence of Kraus operators.
For each Kraus operator of the forward dynamics, $A_{\alpha}$, there should
be a corresponding operator, $\widetilde{A}_{\alpha}$ of the reversed
dynamics such that, starting from equilibrium, the probability of
observing any sequence of Kraus operators in the forward dynamics is the
same as the probability of observing the reversed sequence of reversed
operators in the reversed dynamics. Specifically, for consecutive 
pairs of events, starting from $\pi$ (the  invariant, equilibrium density matrix of the dynamics $\sop{S}\pi = \pi$) 
\[
p(\alpha_1, \alpha_2 | \pi ) = \widetilde{p}(\alpha_2, \alpha_1|\pi)
\]
or equivalently [by Eq.~(\ref{interaction})]
\[
	  \tr \left( A_{\alpha_{2}}A_{\alpha_{1}} \pi 
	  A^{\dagger}_{\alpha_{1}}A^{\dagger}_{\alpha_{2}}
	  \vphantom{\widetilde{A}_{\alpha_{1}}} \right)
	 =
	  \tr \left( \widetilde{A}_{\alpha_{1}}\widetilde{A}_{\alpha_{2}} \pi 
	   \widetilde{A}^{\dagger}_{\alpha_{2}}\widetilde{A}^{\dagger}_{\alpha_{1}}\right) 
\]
Since the invariant density matrix $\pi$, is positive definite it has
a unique inverse and a positive definite square root. We may therefore
insert the identity $I=\pi^{-\frac{1}{2}}\pi^{\frac{1}{2}}$ between
pairs of Kraus operators. By taking advantage of the cyclic property 
of the trace we find that
	\begin{align*}
	  \tr \left( 
	    [\pi^{\frac{1}{2}}A^{\dagger}_{\alpha_{1}}\pi^{-\frac{1}{2}} ]
	    [\pi^{\frac{1}{2}}A^{\dagger}_{\alpha_{2}}\pi^{-\frac{1}{2}}  ]
	      \,\pi\, 
	    [\pi^{-\frac{1}{2}} A_{\alpha_{2}}\pi^{-\frac{1}{2}} ]
	    [\pi^{\frac{1}{2}}A_{\alpha_{1}}\pi^{\frac{1}{2}} ]
	  \vphantom{\widetilde{A}_{\alpha_{1}}} \right)  \\
	  \hfill = \tr \left( 
	    \widetilde{A}_{\alpha_{1}}\widetilde{A}_{\alpha_{2}} 
	      \,\pi\, 
	    \widetilde{A}^{\dagger}_{\alpha_{2}}\widetilde{A}^{\dagger}_{\alpha_{1}}
	  \right) \,.
	\end{align*}
 Therefore, $\widetilde{A}_{\alpha} = \pi^{\frac{1}{2}} 
A^{\dagger}_{\alpha}\pi^{-\frac{1}{2}}$ and the superoperator $\widetilde{\sop{S}}$, the reversal or $\pi$-dual 
of $\sop{S}$, is 
\begin{equation}
	    \widetilde{\sop{S}} \rho= 
	    \sum_{\alpha} 
	      \widetilde{A}_{\alpha} \,\rho\, \widetilde{A}^{\dagger}_{\alpha} =
	    \sum_{\alpha} 
	      [\pi^{-\frac{1}{2}}A^{\dagger}_{\alpha} \pi^{\frac{1}{2}}]
	      \, \rho \, 
	       [\pi^{\frac{1}{2}} A_{\alpha} \pi^{-\frac{1}{2}}]
	  \label{quantumrev0}	       
\end{equation}

If we write $\sop{D}_{\pi}\rho$ for the superoperator 
$\pi^{\frac{1}{2}} \rho \pi^{\frac{1}{2}}$ then this reversal may be 
 expressed independently of any
particular decomposition of $\sop{S}$ into Kraus operators.
\begin{equation}
\boxed{
	    \widetilde{\sop{S}} = \sop{D}_{\pi} \, \sop{S}^{\times} \sop{D}_{\pi}^{-1}
	  \label{quantumrev1}
	  }
\end{equation}
By similar reasoning, the time reversal of the Lindbladian continuous time dynamics [Eq.~(\ref{Lindbladian})] takes the same form, 
$\widetilde{\sop{L}} = \sop{D}_{\pi} \, \sop{L}^{\times} \sop{D}_{\pi}^{-1}$,  analogous to the time reversal of a continuous time Markov chain~\cite{Norris1997}

We can readily confirm that the quantum
operator reversal [Eq.~(\ref{quantumrev1})] is an involution (a duality) on TCP maps with
fixed point $\pi$. From Eq.~\ref{quantumrev0} it is clear that the reversed superoperator has an operator-sum representation and is, therefore, a complete, positive map. Note that $\sop{D}_{\pi}= \sop{D}_{\pi}^{\times}$ is
a Hermitian superoperator (Since positive operators are Hermitian
$\pi^{\dagger}=\pi$), that $\sop{D}_{\pi} I=\pi$, and that $\sop{D}_{\pi} ^{-1}\pi =I$.  Therefore, the reversal is idempotent, 
$\widetilde{\widetilde{\sop{S}}} = \sop{S}$, has the correct invariant
density matrix, 
\[
\widetilde{\sop{S}} \pi = \sop{D}_{\pi}\, 
\sop{S}^{\times}\sop{D}_{\pi}^{-1} \pi = \sop{D}_{\pi}\, 
\sop{S}^{\times} I = \sop{D}_{\pi}\, I = \pi ,
\]
 and is trace preserving.
\[
\widetilde{\sop{S}}^{\times} I =\sop{D}_{\pi}^{-1} \, \sop{S}
\sop{D}_{\pi} I = \sop{D}_{\pi}^{-1} \, \sop{S} \pi = \sop{D}^{-1} \pi = I.
\]
  If $\sop{R}$ and $\sop{S}$ are two TCP maps
with the same fixed point then $\widetilde{\sop{R}\sop{S} }=
\widetilde{\sop{S}}\widetilde{\sop{R}}$.

By analog with classical Markov chain terminology, we may say that 
	a quantum operation is \emph{balanced} with respect to a density matrix, $\pi$, if 
	$\sop{S}\pi = \pi$, and \emph{detailed balanced} if $\widetilde{\sop{S}}
	= \sop{S}$.
Conversely, if $\sop{S}$ is detailed balanced with respect to some 
density matrix $\pi$, then $\pi$ is a fixed point of $\sop{S}$:
\[ 
\sop{S} \pi = \widetilde{\sop{S}} \pi = \sop{D}_{\pi} \sop{S}^{\times} \sop{D}_{\pi} ^{-1} \pi = \sop{D}_{\pi} \sop{S}^{\times} I = \sop{D}_{\pi} I = \pi
\]

In passing, it is interesting to note that the time reversal operation, $\widetilde{\sop{S}} = \sop{D}_{\pi} \, \sop{S}^{\times} \sop{D}_{\pi}^{-1}$ is an anti-linear operator of a superoperator, an anti-super-duper operator.


\section*{Isolated quantum system}
The operator sum representation of a closed system dynamic contains a single, unitary Kraus operator, $U = e^{-\frac{i}{\hbar} H t}$, where $H$ is the system Hamiltonian.
Any density matrix that is diagonal in the energy eigenbasis will be  a fixed point of this dynamics, and any such diagonal operator will commute with the unitary Kraus operator. Therefore, the quantum operator reversal corresponds to the time-reversal of the unitary dynamics.
\begin{align}
\sop{S}(t) \rho &= U \rho U^{\dagger} = e^{-\frac{i}{\hbar}  H t} \; \rho\; 
e^{+\frac{i}{\hbar}  H t} \\
\widetilde{\sop{S}}(t) \rho &= U^{\dagger}  \rho U= e^{+\frac{i}{\hbar}  H t} \;\rho\;
 e^{-\frac{i}{\hbar}  H t} 
\nonumber
\end{align}
	
\section*{Classical Markov chain}
Given an orthonormal basis set $\{\ket{e_i}\}$ we can extract the ``matrix elements'' of a superoperator $\sop{S}$.
\begin{equation}
S_{abcd} = \bra{e_a}\sop{S}(\ket{e_d}\bra{e_c})\ket{e_b}
\end{equation}
There are several different conventions for the ordering of the indices. Caves~\cite{Caves1999} would write $S_{ad,bc}$ and Terhal and DiVincenzo~\cite{Terhal2000}  $S_{ab,dc}$. The ordering used here is convenient when transitioning to a tensor ($\sop{S}^{a\:c\:}_{\:\:b\:d}$) or diagrammatic (${}^{a\rightarrow}_{b\leftarrow} \sop{S}_{\leftarrow c}^{\rightarrow d}$) notation.

In any basis the matrix $M_{ac}= S_{aacc}$  is a Markov stochastic  transition matrix; The elements are real and positive $M_{ac}\geq 0$ and the rows sum to $1$. (Since $\ket{e_c}\bra{e_c}$ is a positive operator and $\sop{S}$ is a positive map, $\sop{S}(\ket{e_c}\bra{e_c})$ must also be a positive operator, and therefore the elements are real and positive. The trace preserving condition requires that $\sop{S}^{\times} I = I$. Since $I= \sum_a \ket{e_a}\bra{e_a}$, therefore $\sum_a S_{aacd} = \delta_{cd}$ and  $\sum_a M_{ac}=1$.)

In the diagonal basis of the equilibrium density matrix, a time-reversal of the quantum operation induces a time reversal of the embedded Markov transition matrix with respect to the probability vector on the density matrix diagonal.
\begin{align*}
\widetilde{M}_{ac}=\widetilde{S}_{aacc} &= \bra{e_a} \pi^{\frac{1}{2}} \sop{S}^{\times}\left(\pi^{-\frac{1}{2}} \ket{e_c} \bra{e_c} \pi^{-\frac{1}{2}}\right) \pi^{\frac{1}{2}}\ket{e_a}
\\
&= \frac{\pi_{aa}}{ \pi_{cc}} \bra{e_a} \sop{S}^{\times}\left( \ket{e_c} \bra{e_c}\right)\ket{e_a}
\\
&=  \frac{\pi_{aa}}{ \pi_{cc}} S_{ccaa} = \frac{p^{\circ}_a}{ p^{\circ}_c} M_{ca}
\end{align*}

\section*{Thermostated quantum system}

The reduced dynamics of a quantum system interacting with an external environment or bath can be derived by considering the deterministic dynamics of the joint system, and then tracing over the bath degrees of freedom, leaving a quantum operation description of the system dynamics alone. 
In particular, this approach provides a concise description of a quantum system interacting with a thermal environment of constant temperature. Let the total Hamiltonian of the combined system be
\begin{equation}
H^{\text{SB}} = H^{\text S} \oplus  I^{\text B}+ I^{\text S}\oplus H^{\text B} + \epsilon H^{\text {int}} ,
\end{equation}
 where $I^{\text S}$ 
and $I^{\text B}$ are system and bath identity operators, $H^{\text S}$ is the Hamiltonian of the system,  $H^{\text B}$ is the bath Hamiltonian, $H^{\text {int}}$ is the bath-system interaction Hamiltonian and  $\epsilon$ is a coupling constant.

We assume that initially the system and bath are uncorrelated, and therefore the initial combined state is  $\rho^{\text S} \oplus \pi^{\text B}$, where  $\pi^{\text B}$ is the thermal equilibrium density matrix of the bath.
\[
\pi^{\text B} = \sum_{i} \frac{e^{-\beta E^{\mathrm B}_{i}}}{Z_{_{\mathrm B}}}
| b_{i}\rangle\langle b_{i} | 	
\]
 Here $\{E^{\text B}_{i}\}$ are the energy eigenvalues, $\{| b_{i}\rangle\}$ are the orthonormal energy eigenvectors of the bath, and $Z_{_{\mathrm B}}$ is the bath partition function.
We follow the dynamics of the combined system for some time, then measure the state of the
bath.
\begin{align}
\sop{S} \rho_{_{\mathrm S}}  &= \tr_{_{\mathrm B}} U_{_{\mathrm{SB}}}
 [\rho_{_{\mathrm S}} \oplus \pi_{_{\mathrm B}}] U_{_{\mathrm{SB}}}^{\dagger} 
 \nonumber \\
&= 
\sum_{j} 
  \langle b_{j} | 
  U_{_{\mathrm{SB}}}
  \left(
  \rho_{_{\mathrm S}}  \oplus
	\left[\sum_{i}\frac{e^{-\beta E^{\mathrm B}_{i}}}{Z_{_{\mathrm B}}}
	| b_{i}\rangle\langle b_{i} | \right] 	
  \right) 
  U^{\dagger}_{_{\mathrm{SB}}}
| b_{j} \rangle
\nonumber \\
&= \sum_{ij} 
\frac{e^{-\beta E^{\mathrm B}_{i}}}{Z_{_{\mathrm B}}}
\langle b_{j} |   U_{_{\mathrm{SB}}} |  b_{i} \rangle	
\;  \rho_{_{\mathrm S}} \;
\langle b_{i} | U^{\dagger}_{_{\mathrm{SB}}} | b_{j} \rangle
\label{thermostated}	
\end{align}
Here, $\tr\subB$ is a partial trace over the bath degrees of freedom, 
$U\subSB= \exp(-i H^{^{\mathrm{SB}}} t /\hbar)$ is the unitary dynamic of the total
system, and we have assumed that the coupling constant $\epsilon$ is small.\
It follows that the Kraus operators for this dynamics are
\begin{equation}
A_{ij} = \frac{e^{-\frac{1}{2}\beta E^{\mathrm B}_{i}}} {\sqrt{Z_{_{\mathrm B}} }} \langle b_{j} | 
U_{_{\mathrm{SB}}} | b_{i} \rangle 
\,.
\label{thermostatedKraus}
\end{equation}
and the corresponding reversed operators are
\begin{align}
\widetilde{A}_{ij} 
&= \pi^{\frac{1}{2}} A^{\dagger}_{ij}\pi^{-\frac{1}{2}} 
\label{reversedKraus}
\\
&= \pi^{\frac{1}{2}} [\frac{e^{-\frac{1}{2}\beta E^{\mathrm B}_{i}}} {\sqrt{Z\subB}}  \langle b_{i} | U^{\dagger}\subSB | b_{j} \rangle]     \pi^{-\frac{1}{2}} 
\nonumber
\\
%
&=\langle b_{i} | [ \sqrtPiS \oplus \sqrtPiB] U^{\dagger}\subSB [\nsqrtPiS \oplus I\subB] | b_{j} \rangle  
\nonumber
\\
&\approx \langle b_{i} | U^{\dagger}\subSB [ I\subS \oplus \sqrtPiB] | b_{j} \rangle  
\nonumber 
\\
&= \frac{e^{-\frac{1}{2}\beta E^{\mathrm B}_{j}}} {\sqrt{Z\subB}}\langle b_{i} | U^{\dagger}\subSB | b_{j} \rangle  \,.
\nonumber
\end{align}
If we compare the reversed operator with the corresponding forward operator  we can see that 
taking the time reversal of a quantum 
operation that acts on the system subspace is equivalent (in the  small coupling limit) to taking the time reversal of the entire system-bath dynamics.

\section*{Driven quantum dynamics}

Consider a system with a time-dependent Hamiltonian, interacting with an external, constant temperature heat bath. We split time into a series of intervals, labeled by the integer $t$. The system Hamiltonian changes from one interval to the next, but is  fixed within each interval. (We can recover a continuously varying system Hamiltonian by making the intervals short.) The dynamics are described by dissipative quantum operations [Eq.~\ref{thermostated}], $\sop{S}_t$ with Kraus operators [Eq.~(\ref{thermostatedKraus})] $\{A^{(t)}_{\alpha}\}$. The dynamical history of the system is defined by the initial system state  $\ket{e_0}\bra{e_0}$, a sequence of interactions between the system and environment, described  by the Kraus operators $A_{\alpha_t}^{(1)}, A_{\alpha_2}^{(2)}, \cdots, A_{\alpha_{\tau}}^{(\tau)}$ and the final system state $\ket{e_\tau}\bra{e_\tau}$.
The probability of observing this history is related to the probability of the reversed history in the time reversed chain
 \begin{equation}
\boxed{
\frac{ p( e_0;\alpha_0, \alpha_1,  \cdots ,\alpha_\tau;  e_\tau ) }
{ \widetilde{p}(e_\tau; \alpha_\tau,\cdots, \alpha_1  ,\alpha_\tau;  e_0)}
=\exp\left\{  -\beta Q   \right\}
}
\label{MR}
 \end{equation}
since for every Kraus interaction of the forward dynamics there is a corresponding interaction in the reverse dynamics such that
$
\widetilde{A}^{\dagger}_{\alpha} = A_{\alpha} \exp\{+\frac{1}{2}\beta Q_{\alpha} \}
$
[Eq.~(\ref{reversedKraus})]
where $Q= -(E_j^{\text{B}} - E_i^{\text{B}})$ is the heat, the flow of energy from the bath to the system during the forward time step.
\begin{align*}
& p( e_0; \alpha_1, \alpha_2,  \cdots ,\alpha_\tau;  e_\tau)  
\\
& = \tr   \bra{e_\tau} A_{\alpha_\tau}^{(\tau)}\cdots A_{\alpha_2}^{(2)} A_{\alpha_1}^{(1)} \ket{e_0}\bra{e_0} A_{\alpha_1}^{(1)\dagger} A_{\alpha_2}^{(2)\dagger} \cdots A_{\alpha_{\tau}}^{(\tau)\dagger} \ket{e_\tau}
\\
&= \tr   
\bra{e_0} \widetilde{A}_{\alpha_1}^{(1)} \widetilde{A}_{\alpha_2}^{(2)} \cdots \widetilde{A}_{\alpha_{\tau}}^{(\tau)} \ket{e_\tau}
\bra{e_\tau} \widetilde{A}_{\alpha_\tau}^{(\tau)\dagger}\cdots \widetilde{A}_{\alpha_2}^{(2)\dagger} \widetilde{A}_{\alpha_1}^{(1)\dagger} \ket{e_0}
\\
&\qquad \times \exp\left\{  -\beta ( Q_{\alpha_1} + Q_{\alpha_2}  +\cdots+  Q_{\alpha_\tau} )   \right\} 
\\
&= { \widetilde{p}(e_\tau; \alpha_\tau,\cdots, \alpha_1  ,\alpha_\tau;  e_0)}
\exp\left\{  -\beta Q  \right\} 
\end{align*}

Recall that $Q$ is the heat flow into the system, $\beta$ is the inverse temperature of the environment, and therefore $-\beta Q$ is the change in entropy of that environment. 
This property of microscopic reversibility as expressed in Eq.~(\ref{MR})~\cite{Crooks1998}, immediately implies that the work fluctuation relation~\cite{Crooks1999a} 
and Jarzynski identity~\cite{Jarzynski1997a} 
can be applied  to a driven quantum system coupled to a thermal environment~\cite{Kurchan2001,Mukamel2003a, DeRoeck2004,Talkner2007b,Crooks2007d}. 
The crucial difference between the classical and quantum regimes is that in the quantum case we must avoid explicitly measuring the work directly, which is tantamount to continuously monitoring the system, and instead measure the heat flow from the environment~\cite{DeRoeck2004,Crooks2007d}. 

Financial support was provided by the DOE/Sloan Postdoctoral Fellowship in Computational Biology; by the Office of Science, Biological and Environmental Research, U.S. Department of Energy under Contract No. DE-AC02-05CH11231; and by California Unemployment Insurance.

\bibliography{GECLibrary}	

\end{document}